\documentclass[twocolumn,a4paper]{article}

\usepackage[sc]{mathpazo} 
\usepackage[T1]{fontenc} 
\usepackage[utf8]{inputenc}
\usepackage{microtype} 

\usepackage[hmarginratio=1:1,top=32mm,columnsep=20pt]{geometry} 
\usepackage{multicol} 
\usepackage[hang, small,labelfont=bf,up,textfont=it,up]{caption} 
\usepackage{booktabs} 
\usepackage{float} 
\usepackage{graphicx}
\usepackage[colorlinks]{hyperref} 

\usepackage{paralist} 

\usepackage{abstract} 

\graphicspath{{Images/}{./}}

\usepackage{titlesec} 
\renewcommand\thesection{\Roman{section}} 
\renewcommand\thesubsection{\Roman{subsection}} 
\titleformat{\section}[block]{\large\scshape\centering}{\thesection.}{1em}{} 
\titleformat{\subsection}[block]{\large}{\thesubsection.}{1em}{} 

\usepackage{fancyhdr} 
\title{Feature Extraction from Segmentations of Neuromuscular Junctions\footnotemark} 

\author{
\large
\textsc{Julia Portl}\\ 
\normalsize Interdisciplinary Center for Scientic Computing, Heidelberg University \\ 
\normalsize \href{mailto:julia.portl@iwr.uni-heidelberg.de}{julia.portl@iwr.uni-heidelberg.de}
\\[2mm] 
\textsc{Heike Leitte}\\ 
\normalsize Technical University of Kaiserslautern \\
\normalsize \href{mailto:heike.leitte@iwr.uni-heidelberg.de}{heike.leitte@cs.uni-kl.de} 
\vspace{-5mm}
}
\date{}

\begin{document}
    
\twocolumn[
 \begin{@twocolumnfalse}
   \maketitle
   \begin{abstract}
       \noindent
       Segmentations are often necessary for the analysis of image data. They are used to identify different objects, for example cell nuclei, mitochondria, or complete cells in microscopic images. There might be features in the data, that cannot be detected by segmentation approaches directly, because they are not characterized by their texture of boundaries, which are properties most segmentation techniques rely on, but morphologically. 
       In this report we will introduce our algorithm for the extraction of suchlike morphological features of segmented objects from segmentations of neuromuscular junctions and its interface for informed parameter tuning.
   \end{abstract}
 \end{@twocolumnfalse}
]


\section{Introduction}
\footnotetext{This report is based on a part of a former paper submission in 2012.
    A shortened version of this submission \cite{portl13} was published in 2013 and can be consulted for more information about the biological background. It skipped technical details of the feature extraction and focuses on the further analysis procedure of
    these features.}

\begin{figure*}[htbp]
     \centering
     \includegraphics[width=.9\linewidth]{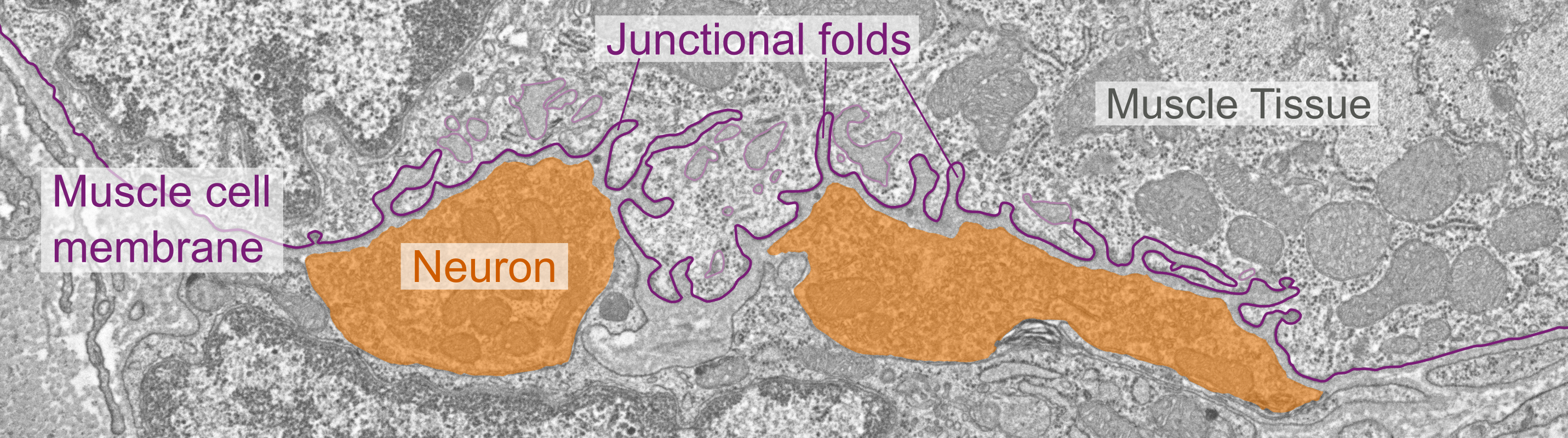}
     \caption{\label{fig:segmentedData}
        Microscopy image of a neuromuscular junction, the interface between a muscle cell and its neuron. The muscle membrane features the typical folds where it is located close to the neuron. Our goal is to find the points on the membrane where a fold begins and ends, respectively. }
    \end{figure*}

In our application, biologists are interested in junctional folds and their characteristics. Junctional folds are parts of the muscle cell's membrane that dig finger-like into the surrounding muscle tissue (see Figure \ref{fig:segmentedData}). 
 In the image data, the membrane is represented by a single line-like structure. The segmentation of this structure is given by a polygonal chain, which does not distinguish between individual folds. Hence, we have to extract the fold from the polygonal chain in order to compute fold characteristics such as width, depth, or branching style.

\section{Definitions}
\label{subsec:infoex}
We define a \emph{fold} as a (connected) subchain of a polygonal chain, satisfying two conditions:
\begin{enumerate}
    \item The subchain contains a turning angle greater than a given threshold $\tau$. A subchain that satisfies this condition is
    called a \emph{minimal subset}.
    \item The first and the last point of the subchain have a maximum Euclidean distance of $2\delta$ and the arc length of the subchain is \mbox{$\gg\!2\delta$}.
    A subchain that satisfies this condition is called a \emph{maximal subset}. 
\end{enumerate}
Condition 1 specifies that we are looking for parts of the membrane that feature a tip and condition 2 requires the structures to have a small baseline, i.e., that the membrane has a close distance to itself at the end of the fold.

For the extraction of subchains that satisfy the first condition we use a simplified version of an orientation based shape matching algorithm introduced in \cite{CohenGuibas97}. 

Let $P_n$ be a piecewise linear curve (poly\-gonal chain, polyline) given by $n+1$ vertices $\mathbf{v}^0, \mathbf{v}^1, \dots, \mathbf{v}^{n} \in \mathbb{R}^2$.
Let $\mathbf{s}^i = \mathbf{v}^{i+1} - \mathbf{v}^i$ be a segment of $P_n$ and $l_i$ be the arc length of a curve $P_i, \; i=0,1,\dots,n$.
The parametric representation $P_n(t)$ of $P_n$ is given by

\begin{equation}
    P_n(t) = \frac{ ( t-l_i )  \mathbf{s}^i }
    {|\mathbf{s}^i|} + \mathbf{v}^i,
\end{equation}
with $t \in [l_i, l_{i+1} ), \; i = 0,1,\dots,n$.

The orientation function $O_{P_n}(t)$ of $P_n$ is piecewise constant and defined as
\begin{equation}
    O_{P_n}(t) = \arctan\!2\,( \mathbf{s}^i_y, \mathbf{s}^i_x ),
\end{equation}
with $t \in [l_i, l_{i+1} ), \; i = 0,1, \dots, n$.

$O_{P_n}(t)$ is smoothed using a low-pass filter on its Fourier transform to eliminate small irregularities in membrane.
Let $\tilde{O}_{P_n}(t)$ be the smoothed orientation function.

\section{Minimal subsets}
Minimal subsets, highlighted in light blue in Figure \ref{fig:foldDetection_minMax}, are extracted from the smoothed orientation function $\tilde{O}_{P_n}(t)$ by means of extreme value analysis. 
All folds on a polyline are either oriented to its left or to its right. 
Folds to the left appear in the smoothed orientation function as maximum/minimum pairs, folds to the right as minimum/maximum pairs.
Hence, the order of the succession of minima and maxima within an extrema pair (min/max vs. max/min) is consistent within a polygonal chain. 

For each consecutive local maximum/minimum (or minimum/maximum) value pair, that satisfies the first condition, we require a minimal distance between these extreme values greater than $\tau$. We found $\tau = 2/3 \pi$ to be a good default parameter that gave fair results in most of our data sets. 
The minimal subsets comprise regions on the polyline where it describes a curve and parts of the polyline are oriented face-to-face.


\section{Maximal subsets}
\begin{figure*}[htbp]
    \centering
    \includegraphics[width=.75\linewidth]{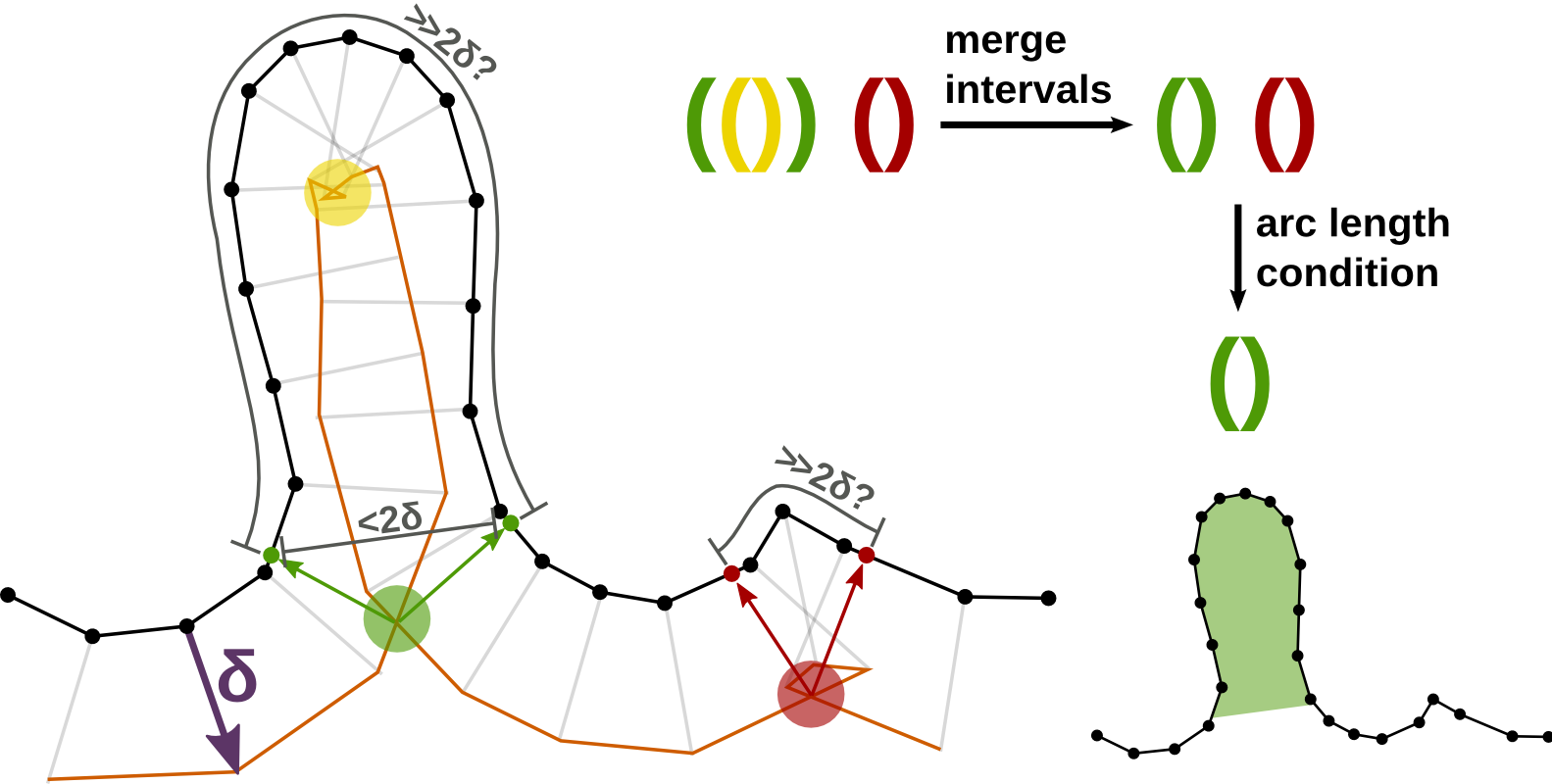}
    \caption{\label{fig:maximalSubsets}
        Procedure of obtaining maximal subsets. The orange polyline results from offsetting the original membrane line by the value $\delta$. Its self intersections (marked in yellow, green, red) are translated into intervals and overlapping intervals are merged. Intervals whose arc length along the membrane are shorter than the Euclidean distance between their end points are discarded. }
\end{figure*}

Condition 2 can be fulfilled with the aid of self-intersections of \emph{polyline offsets}. Their detection is illustrated in Figure \ref{fig:maximalSubsets}: First, we apply an offset $\delta$ to each point in the membrane polyline in the direction of the weighted average of normals from adjacent segments. This offset line (orange) features self-intersections denoting points on the polyline, whose distance is smaller than $2\delta$. The intersection points are back-projected onto the polygonal chain indicating limits of membrane intervals fulfilling condition 2.
There are usually multiple intersections of the offset lines within one fold. 
Therefore, we unite overlapping intervals and keep the one that yields in the longest subchain of the membrane with end points being less than $2\delta$ apart. A good default value for $\delta$ is the width of an average fold.

Intervals whose arc length along the membrane are shorter than the Euclidean distance between their end points are discarded.

\section{Interactive parameter tuning}
Three parameters have to be set for the feature extraction; two for the determining minimal subsets and one for the maximal subsets. 
We propose default values for our parameters that usually work very well on most data sets.
Assuming that a data set originates from the same tissue sample and passed through the same segmentation process, it is in general sufficient, if the parameters are tuned once for a small part of the data set, e.g., for a single slice of a data volume, 
and then applied to the rest of the data. 
Figure \ref{fig:foldDetection_minMax} displays what is shown to the user when tuning these three parameters.
We display the orientation function, since folds are detected using this function, to make the detection algorithm and the effect of single parameters more transparent to the user.
This allows for an informed and intuitive adjustment of parameters.

\subsection{Smoothing factor and turning angle}
The minimal subsets depend on the turning angle difference $\tau$ and on the smoothing factor of the orientation function.
The latter again depends on the initial roughness of this function. 
While changing the smoothing factor, the user can observe the original function overlaid with the smoothed function and its local extrema (see Figure \ref{fig:foldDetection_minMax} right). Thus, she can easily choose a suitable value.
Manipulation of the turning angle parameter $\tau$ is usually not necessary. 
However, a decrease results in the disappearance of complete minimal subsets where the vertical distance of a minima/maxima pair falls below the threshold $\tau$ in the smoothed orientation graph. 
The minimal subsets are displayed as intervals represented by blue boxes. These are plotted above the orientation function using the same x-axis. This clarifies their correspondence to the orientation function and its extreme values. 
Additionally, the intervals are highlighted in blue on the original pink polygonal chain (Figure \ref{fig:foldDetection_minMax} left).

\subsection{Polyline offset}
The maximal subsets are depicted as intervals represented by green boxes above the orientation function, sharing the same x-axis.
The influence of tweaking the offset $\delta$ can directly be observed in the resulting offset polyline (left: green polyline parallel to the pink one) and the areas highlighted in green on the original polyline.
A higher offset value results in wider intervals, i.e.\ folds "grow" more along their sides. Since there is no exact definition of where a fold starts, this choice should be left to the domain expert. 
With respect to the further analysis of the folds, it is necessary that the offset $\delta$ stays fixed for the whole data.
A good default value for $\delta$ is the width of folds in the data set, which is more or less the same within one tissue sample.

\section{Junctional folds}
\begin{figure*}[htbp]
    \centering
    \includegraphics[width=.69\linewidth]{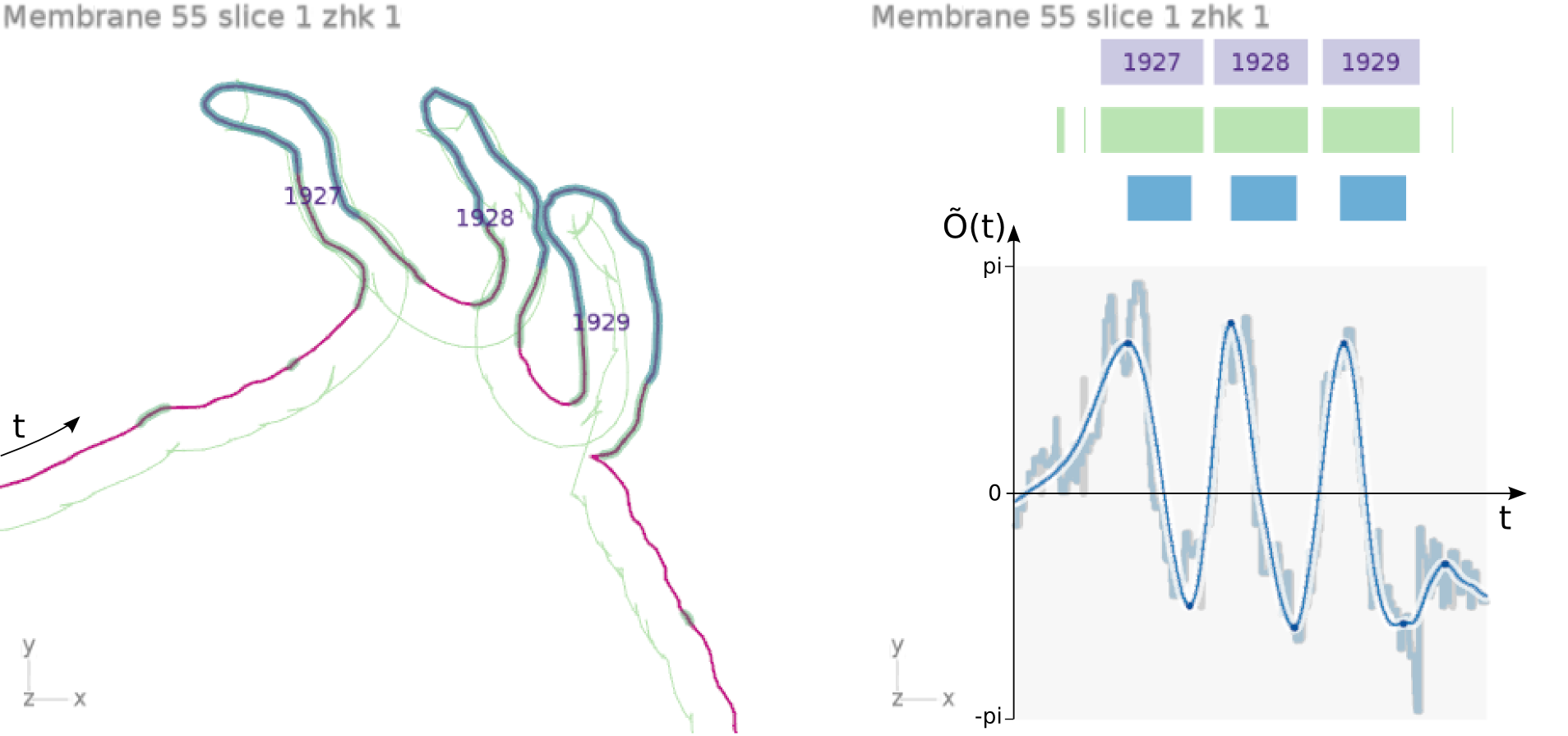}
    \caption{\label{fig:foldDetection_minMax}
        The three descending slopes of the orientation function on the right hand side correspond three \emph{minimal subsets}. The
        intersections of the green polyline on the left determine the \emph{maximal subsets}. The intervals are represented in
        blue (minimal) and green (maximal) as colored blocks above the function and on the polyline on the left. }
\end{figure*}

Junctional folds can now be identified by checking for each \emph{maximal subset} if it also contains a \emph{minimal subset}. 
Figure \ref{fig:foldDetection_minMax} shows a small detail of a neuromuscular junction, a membrane part that features three folds. The final folds are labeled with numbers and their interval representation is depicted in violet.

Note that, in general, using maximal subsets exclusively are insufficient to determine folds.
This approach tends to overestimate the existence of a fold, which can be observed in Figure \ref{fig:foldDetection_minMax}, top right:
Six maximal subsets (green blocks) are found in this example, whereas minimal subsets (blue blocks) have been detected at three positions. 
Only the combination of both criteria (violet blocks) gives the desired results.

\section{Results}
\begin{figure}[htbp]
    \centering
    \includegraphics[width=.8\linewidth]{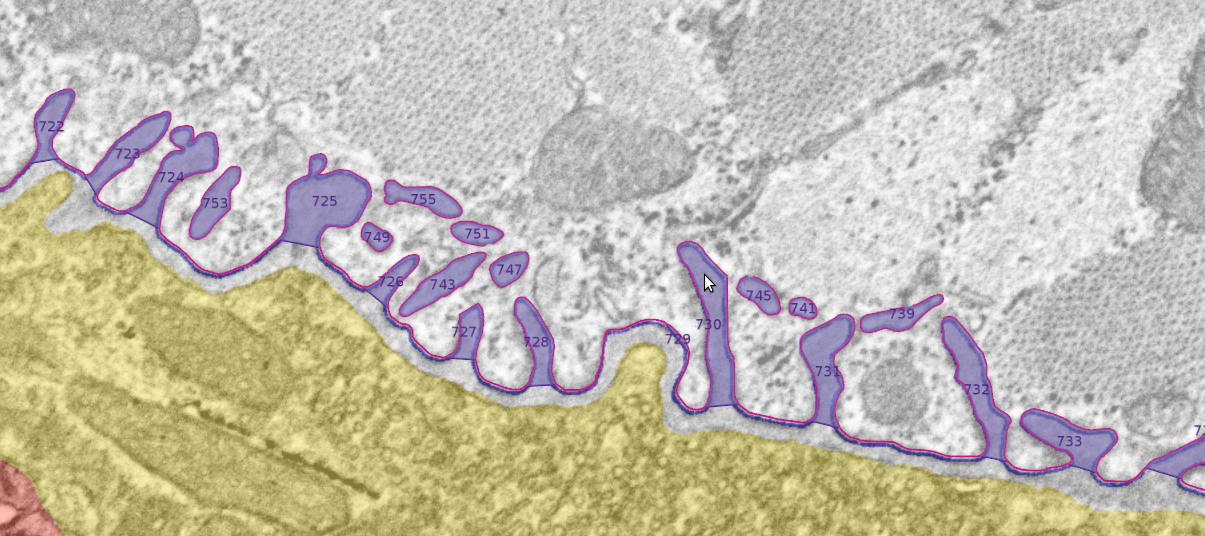}
    \caption{\label{fig:foldDetection_combined}
        Detected folds (violet) on a muscle cell membrane (magenta). A large variety of shapes is covered by the extraction.}
\end{figure}

For most cases the default values work very well for the detection of folds in a polygonal chain and even detected very unusual folds.
Figure \ref{fig:foldDetection_combined} shows an example of the proposed fold extraction technique on a part of a neuromuscular junction. All folds were automatically detected. The bulge of the neuron (yellow) in the center was correctly considered not to be a fold by our algorithm.
If required, the parameters can be adjusted interactively and the result can be judged instantly.
The resulting folds are projected onto the microscopy images, so that the user can judge the correctness of their extraction in their biological context.

\section{Conclusion}
Our algorithm detects single folds-like features on polylines from segmentations of muscle cell membranes of neuromuscular junctions.
We combined two criteria for defining a fold and used these for the extraction the folds.
We provide a strategy for interactive, intuitive and informed parameter tuning by visualizing impacts of the parameters on details of the detection algorithm.


\bibliography{egbibsample}{}
\bibliographystyle{eg-alpha}


\end{document}